\documentstyle[12pt]{article}
\begin{document}
\makeatother
\renewcommand{\theequation}{\thesection.\arabic{equation}}
\newcommand{\am}{a \times m}
\newcommand{\ri}{RI }
\newcommand{\mathleftline}[1]{\hbox to0pt{\hss\hbox to12pt{\hbox
 to\hsize{$#1$\hfill}\hss}}}
\newcommand{\COM}[1]{[[[#1]]]}
\newcommand{\ot}{\frac{1}{2}}
\newcommand{\D}{& \displaystyle}  
\newcommand{\di}{\displaystyle}   

\date{June 1993}
\title
{Finite-Size Scaling in the $O(n)$ $\phi^4_4$ Model\thanks{Supported by
Fonds zur F\"orderung der Wissenschaftlichen Forschung in
\"Osterreich, project P7849.} }
\author
{\bf R. Kenna \\  \\
Institut f\"ur Theoretische Physik,\\
Universit\"at Graz, A-8010 Graz, AUSTRIA}
\maketitle
\begin{abstract}
Perturbation theory and renormalization group methods are  used  to
derive   a  finite-size  scaling  theory for the partition function
zeroes and  thermodynamic functions in the $O(n)$ $\phi^4$ model in
four dimensions.  The  leading  power--law scaling behaviour is the
same  as  that  of  the  mean  field  theory. There exist, however,
multiplicative logarithmic corrections  which  are  linked  to  the
triviality of the theory.

\end{abstract}

\vspace*{-15cm}
\hfill \parbox{3cm}{ UNIGRAZ-- \\
                     UTP-- \\
                     28--06--93 \\
               ~ }\\
\vspace*{2cm}

\newpage

\section{Introduction}
\setcounter{equation}{0}

The construction of the  continuum limit  of  a quantum field theory
regularized on a Euclidean space--time lattice  is equivalent to the
study  of  the  critical  behaviour  of  certain statistical physics
models.   While   in   the   quantum   field   theory   context  the
$O(4)$--symmetric $\phi^4$ theory (in four dimensions) appears as an
essential  part  of  the  Higgs   sector  of  the standard model, in
statistical physics  (below  four  dimensions)  the  $O(n)$   theory
provides  a  model  of  ferromagnetism.  Above  the  upper  critical
dimension $d=4$, the scaling behaviour  simplifies and  the critical
exponents  are  exactly those given  by the mean field theory. It is
rigorously  known  that  the  continuum  limit  is  then trivial and
described by free (Gaussian) fields  \cite{triviality}.  Below  four
dimensions     the   super--renormalizable  $\phi^4_d$   theory   is
non--trivial \cite{nontriviality}.

The   action   for  the  $n$--component  theory  in  $d$--dimensional
Euclidian space--time continuum is
\begin{equation}
 S
 =
 \int d^dx
 \left\{
  \frac{1}{2}(\nabla {\vec{\phi}})^2 + \frac{m_0^2}{2}{\vec{\phi}}^2
            + \frac{g_0}{4!}({\vec{\phi}}^2)^2
 \right\}
\quad ,
\label{action}
\end{equation}
where  $\vec{\phi}$ is a vector in the $n$--dimensional internal space,
$g_0$  the bare quartic self coupling and $m_0$   the bare bosonic mass
in the quantum field theory.

Whereas in statistical physics the bare quantities have direct physical
meaning, in field theory it is the renormalized quantities which are of
interest. Because  there is no exact solution for the $O(n)$ $\phi^4_4$
theory approximations such as perturbation theory (in  the renormalized
mass  and  the quartic  self  coupling), high  and low  temperature (or
coupling) expansions or Monte Carlo (MC) methods have   to  be used. In
\cite{BLZ}    the  renormalization  group (RG)    was   combined   with
perturbative  methods   to analyse  the approach of the infinite volume
theory  to  criticality.  If ${m_0}^2$  is written as  ${m_0}_c^2 + t$,
where ${m_0}_c$ is the critical bare    mass for which the renormalized
theory is massless, then $t$ is a measure   of the distance        from
criticality.     In the statistical physics analogue   the squared mass
corresponds to the Boltzmann factor and $t$ is called       the reduced
temperature. The (perturbative) predictions of \cite{BLZ} for the  four
dimensional      scaling     behaviour   of   the  correlation function
$\int d^4x \langle \phi(x)\phi(0) \rangle$    and   the  energy--energy
correlation $\int d^4 x \langle \phi^2(x)\phi^2(0) \rangle$        (the
susceptibility  and  specific  heat  respectively  in  the  statistical
mechanics analogue) are
\begin{equation}
 \chi_\infty (t)  \sim  c_1 |t|^{-1}|\ln{|t|}|^{\frac{n+2}{n+8}}
\quad ,
\label{two}
\end{equation}
and
\begin{equation}
  C_\infty (t)  \sim   c_2  |\ln{|t|}|^{\frac{4-n}{n+8}}
  + {\rm{const.}}
\label{three}
\end{equation}
for $n \neq 4$. If $n=4$ (\ref{three}) becomes
\begin{equation}
  C_\infty (t) \sim
  c_3 |\ln{|\ln{|t|}|}|   + {\rm{const.}}
\quad .
\label{three'}
\end{equation}
In the above formulae the $c_j$ have the form
\begin{equation}
 c_j = {\rm{const.}}
 \left\{ 1 + O \left(
                   \frac{\ln{|\ln{|t|}|}}{\ln{|t|}}
                     \right)\right\}
\quad .
\label{cjcj}
\end{equation}
The correlation length in four dimensions is \cite{LW}
\begin{equation}
 \xi_\infty (t) \sim |t|^{-\frac{1}{2}}
 \left|
  \ln{|t|}
 \right|^{\frac{n+2}{2(n+8)}}
\quad .
\label{four}
\end{equation}
The leading power law scaling behaviour of these thermodynamic functions
in four dimensions coincides with the predictions of mean field  theory.
In the formulae (\ref{two}) to (\ref{four}) there are     multiplicative
logarithmic corrections      which         are intimately  linked to the
triviality of the theory. Indeed, it has been rigorously  proven that if
any multiplicative logarithmic corrections to the (mean   field) scaling
behaviour of the susceptibility are present in a continuum   limit which
is used to construct  a  $\phi_4^4$  field  theory, this limiting theory
must  be  trivial  \cite{AiGr83}.   This  has  been   shown  to  be  the
case ((\ref{two})  and (\ref{four}) have been            proven) for the
weakly coupled version of the single component theory (i.e.,   for $g_0$
small  enough)   \cite{HaTa87}.    In   the   strong   coupling   limit,
$g_0 \rightarrow \infty$,   the   $O(n)$   $\phi^4$   model  becomes the
$n$--vector  model   (also   called  the   non--linear $\sigma$  model).
It is believed  that this theory  remains in the same universality class
as the weakly coupled version. It is therefore   important   to    check
the above formulae in  a non--perturbative  fashion.  Such  an  approach
is the MC method  which  involves   the  use of stochastic techniques to
calculate path integrals.

While  for  $n\geq 1$  computational   resources limit one to relatively
small  lattices,  the $n=0$ case introduces significant simplifications.
The $n=0$ limit of the $n$--vector model is the self avoiding random walk
problem.     In \cite{ADCCaFr} these simplifications were used to verify
(\ref{two}) and (\ref{four}) numerically. A numerical analysis along the
lines of \cite{ADCCaFr} is not feasible        for the $n\geq 1$ theory.
Instead a finite--size extrapolation  theory   is   needed. Finite--size
scaling   (FSS)   was   first   formulated   by Fisher   and co--workers
\cite{FiFe67,Fi72}.   This  allows  one  to  extract  information on the
critical  behaviour  of  systems  in  the  thermodynamic  limit  from an
analysis  of  the  (finite)   size  dependency  of certain thermodynamic
quantities .

A finite--size scaling theory for the partition function zeroes   of the
single   component   version  of  the $\phi^4_4$ model has recently been
derived and found   to  be  in  quantitative  agreement with a numerical
analysis of the four dimensional Ising model \cite{KeLa93}.

The main purpose  of the present work is to extend the FSS theory     of
\cite{KeLa93}  to the $n$--component $\phi_4^4$ model.   The strategy of
\cite{KeLa93}, in which FSS is first established   for the zeroes of the
partition function and consequently for the  thermodynamic functions, is
followed here.

The layout of this paper is as follows. In sect.2 the FSS hypothesis and
the  concept  of   partition   function   zeroes   are   recalled.   The
(perturbative) renormalization   group   equation  (RGE)  for  the  free
energy  of  a finite--size  system  is solved in sect.3. This is used to
find  the  FSS  of  the partition  function zeroes and the thermodynamic
functions in sect.4 and conclusions are drawn in sect.5.

\section{Finite--Size Scaling and Partition Function Zeroes}
\setcounter{equation}{0}

The usual statement of FSS for any thermodynamic quantity which exhibits
an algebraic singularity in the infinite system    is that near the bulk
critical point $t=0$ \cite{Ba83}
\begin{equation}
  \frac{P_l(t)}{P_\infty(t)}
  =
  f\left(
   \frac{l}{\xi_\infty (t)}
  \right)
\quad .
\label{FSShyp}
\end{equation}
Here the $l$ denotes the linear extent of the system, $\xi_\infty (t)$
is the correlation length of the infinite size system and  $f$  is  an
unknown function of its argument which is called the scaling variable.

FSS in  its above form  was derived from RG considerations by Br\'ezin
below  four dimensions  \cite{Br82}.      The derivation relies on two
assumptions over   and above the usual assumptions of RG theory. These
are  that  the  system  length  $l$ is  not  renormalized  in the flow
equations  and  that the infra--red fixed point (IRFP) $g_R^*$  is not
zero. This  latter  assumption fails in four dimensions.

The continuum parameterization of a quantum field theory is recovered
from  its lattice regularized version at a phase transition of second
order. In 1952 Lee and Yang provided an alternative way to understand
the onset of a phase transition. The behaviour of  all  thermodynamic
quantities  is  determined  by  the (complex) zeroes of the partition
function. In the complex plane of the external odd ordering field the
Lee--Yang  theorem  states  that  all  of  these  zeroes  lie  on the
imaginary axis  (when  the  remaining parameters governing the system
are real) \cite{LY}.  These  zeroes   are  called Lee--Yang zeroes to
distinguish them  from  the  zeroes in the reduced temperature plane.
The study of the latter type of zeroes was first emphasized by Fisher
\cite{Fi72} and they are  referred to as temperature zeroes or Fisher
zeroes. For a finite system,   for which the partition function is an
analytic function of its parameters, there is no phase transition and
the zeroes are truly   complex (i.e., non--real). A phase transition,
which manifests itself as a point of non--analycity can only arise if
the zeroes pinch the real axis in the thermodynamic  limit.

Itzykson,  Pearson  and  Zuber  managed  to  connect  the concepts of
partition  function  zeroes  and the RG thereby deriving a FSS theory
for  the  former  below  four  dimensions  \cite{IPZ}.  A finite size
scaling  theory  for  the  single  component   $\phi^4_4$  theory was
developed in \cite{KeLa93} where it was also shown how to extract the
FSS of thermodynamic quantities from the scaling of the Lee--Yang and
Fisher zeroes. The primary purpose of the present  work is to present
the FSS of the $O(n)$ version (in four dimensions).   To this end the
FSS behaviour of the partition function zeroes,   the    zero   field
magnetic  susceptibility and the specific heat are calculated. All of
these results are also obtainable from the      modified finite--size
scaling hypothesis proposed in \cite{KeLa93}.

\section{The Renormalization Group Equations}
\setcounter{equation}{0}

The  vacuum  to  vacuum  transition  amplitude for the $n$--component
$\phi_d^4$   theory  in   the    presence   of   an   external source
$\vec{H}(x)$  represents the evolution of the vacuum via the creation
interaction and  destruction of   particles through the medium of the
source. In Euclidean space--time       continuum, and in terms of the
generating functional for the connected Green's functions it is
\begin{equation}
  \exp{W[{\vec{H}},t]}
  \propto
  \int \prod_x \prod_\alpha d \phi_\alpha (x)
  \exp{\left( -S \right)}
\quad ,
\label{thesis2.7}
\end{equation}
where the constant of proportionality  is such that $W[\vec{0},0] = 0$,
and $\alpha = 1,\dots,n$ label the field components. The action is
\begin{equation}
 S
 =
 \int d^dx
 \left\{
  \frac{1}{2}(\nabla {\vec{\phi}})^2 + \frac{m_0^2}{2}{\vec{\phi}}^2
            + \frac{g_0}{4!}({\vec{\phi}}^2)^2
            - \frac{t(x)}{2} {\vec{\phi}}^2(x)
            + {\vec{H}}(x) {\vec{\phi}}(x)
 \right\}
\quad ,
\label{thesis2.6}
\end{equation}
where $t(x)$ is a source for quadratic composite fields. Its inclusion
facilitates the derivation of energy--energy correlation functions.

The function conjugate to ${\vec{H}}$ is
\begin{equation}
 M_\alpha = \frac{\delta W[\vec{H},t]}{\delta H_\alpha(x)}
\quad .
\label{thesis2.9}
\end{equation}
The generating functional  for the  one particle irreducible (1PI) vertex
functions (or Schwinger functions) is written as $\Gamma [{\vec{M}},t]$,
and is given by a following Legendre transformation on $W[{\vec{H}},t]$;
\begin{equation}
\Gamma [{\vec{M}},t] + W[{\vec{H}},t]
 = \int dx {\vec{H}}(x){\vec{M}}(x)
\quad .
\label{thesis2.10}
\end{equation}
This gives
\begin{equation}
 H_\alpha[x,t]
 =
 \frac{\delta \Gamma [\vec{M},t]}{\delta M_\alpha (x)}
\quad .
\label{thesis2.11}
\end{equation}

No   difficulties   beyond   those  already encountered in the single
component  $\phi_4^4$   theory  studied in \cite{KeLa93} arise in the
$n$--component    version.    The critical theory (given by the value
${m_0^2}_c$ of $m^2_0$ for which  the renormalized theory is massless)
is firstly renormalized.      The bulk renormalization  constants are
sufficient   to   renormalize    the   finite--size   theory and this
renormalization is performed at an arbitrary non--zero mass parameter
$\mu$ in order to control infra--red divergences.  In four dimensions
the vertex function having no external legs and two  composite fields
has to be additively renormalized and this  gives    rise    to    an
inhomogeneous   term  in the RGE. One then applies a Taylor expansion
(in $t$ and ${\vec{M}}$)  to the vertex functions to find the RGE for
the    massive   theory   (in   the   critical   region).   Following
\cite{BLZ,KeLa93} the solution for the free energy is (after applying
dimensional analysis)
\begin{eqnarray}
\lefteqn{
       \Gamma_R^{(0,0)}(t,\vec{M},g_R,\mu,l)
}
\nonumber \\
  &  &
       =
       l^{-4}
       \Gamma_R^{(0,0)}(l^2t(\lambda),l\vec{M}(\lambda),
        g_R(\lambda),l\mu(\lambda),1)
       + \Pi_n(\lambda;t)
\quad  ,
\label{thesis5.2}
\end{eqnarray}
where
\begin{equation}
 \Pi_n(\lambda;t) =       -
       \frac{1}{2!}
       \int_1^\lambda
           \frac{d\lambda'}{\lambda'}
           {t(\lambda')}^2
           \Upsilon(g_R(\lambda^\prime))
\end{equation}
Here, $\mu (\lambda) = \lambda \mu$ is a rescaling of the arbitrary mass
$\mu$. The functions $g_R(\lambda)$, $M_\beta (\lambda)$,   $t(\lambda)$
and $\Upsilon (\lambda)$ respond to this rescaling through the      flow
equations.  To leading order in perturbation theory these flow equations
for  the $O(n)$ theory are \cite{BLZ}
\begin{eqnarray}
  \frac{ d g_R (\lambda) }{ d \ln{\lambda} }
  & = &
  \frac{n+8}{6}
  {g_R(\lambda)}^2
  \left\{ 1+ O(g_R ( \lambda )) \right\}
\quad ,
\label{thesis5.69}
\\
  \frac{d \ln{M_\alpha}(\lambda)}{d \ln{\lambda}}
 & = & - \frac{n+2}{144} g_R(\lambda)^2
    \left\{ 1+ O(g_R (\lambda)) \right\}
\quad ,
\label{thesis5.70}
\\
 \frac{d \ln{t(\lambda)}}{d \ln{\lambda}}
  & = &  \frac{n+2}{6} g_R(\lambda)
   \left\{ 1+ O({g_R (\lambda)}^2) \right\}
\quad ,
\label{thesis5.71}
\\
  \Upsilon (g_R(\lambda)) & = & \frac{n}{2}\left\{ 1
        + O(g_R (\lambda)) \right\}
\quad .
\label{thesis5.72}
\end{eqnarray}
For $\lambda \ll 1$ the solutions to these flow equations are
\begin{eqnarray}
 g_R(\lambda ) & = & a_1 (-\ln{\lambda})^{-1}
\quad ,
 \\
 t (\lambda )  & = & a_2  t (-\ln{\lambda})^{-\frac{n+2}{n+8}}
\quad ,
 \\
 M_\alpha (\lambda ) & = & b_1 M_\alpha
\quad ,
\end{eqnarray}
\begin{equation}
  \Pi_n(\lambda;t) \propto   \left\{
                           \begin{array}{ll}
  t^2 \left[
             a_3 (-\ln{\lambda})^{\frac{4-n}{n+8}}
             +
             {\rm{const.}}
      \right]
              &  {\rm for\ } n \neq 4
                                                                \\
      t^2 \left[
                 b_2 \ln{(-\ln{\lambda})}  + {\rm{const.}}
            \right]
                                         & {\rm for\ }  n = 4
               \quad ,
               \end{array}
          \right.
\label{thesis5.84'}
\end{equation}
where  the $a_j$ have the form
\begin{equation}
  {\rm{const.}} \left\{ 1 +
        O\left( \frac{\ln{|\ln{\lambda}|}}{\ln{\lambda}}
  \right)  \right\}
\label{1stc}
\end{equation}
and the $b_j$ are
\begin{equation}
 {\rm{const.}}
 \left\{
          1 + O\left(
                     \frac{1}{\ln{\lambda}}
               \right)
 \right\}
\quad .
\label{2ndc}
\end{equation}
Now, choosing $\lambda = l^{-1}$ and applying perturbation theory to the
homogeneous  term  of (\ref{thesis5.2})  as  in \cite{KeLa93}, one finds
\begin{eqnarray}
\lefteqn{
         \Gamma_R^{(0,0)}\left( t,\vec{M},g_R,1,l  \right)
}
\nonumber \\
& &
 = a_1^\prime t{\vec{M}}^2 \left( \ln{l}  \right)^{-\frac{n+2}{n+8}}
   +
   a_2^\prime  ({\vec{M}}^2)^2 (\ln{l})^{-1}
   +
   \Pi_n(l^{-1};t)
\quad ,
\label{thesis5.99}
\end{eqnarray}
where $a_1^\prime$ and $a_2^\prime$ are of the form (\ref{1stc})   above
with $\lambda$ replaced by $l^{-1}$. Applying (\ref{thesis2.11}) to this
yields for the external field
\begin{equation}
  H_\alpha \left( t,{\vec{M}},g_R,1,l  \right)
 =
 2 a_1^\prime t M_\alpha (\ln{l})^{-\frac{n+2}{n+8}}
 +
 4 a_2^\prime (\vec{M}^2) M_\alpha (\ln{l})^{-1}
\quad .
\label{thesis5.101}
\end{equation}
The free energy per unit volume in the presence of an external field  is
\begin{equation}
  W_l(t,\vec{H}) = \vec{M} \vec{H}(t,\vec{M};l)
  -
  \Gamma_R^{(0,0)}(t,\vec{M};l)
\quad .
\label{thesis5.102}
\end{equation}
{}From (\ref{thesis5.99}) and (\ref{thesis5.101}) this is
\begin{equation}
 W_l(t,\vec{H})
 = a_1^\prime  t {\vec{M}}^2 (\ln{l})^{-\frac{n+2}{n+8}}
   +
   3 a_2^\prime  ({\vec{M}}^2)^2   (\ln{l})^{-1}
   -
   \Pi_n(l^{-1};t)
\quad .
\label{thesis5.103}
\end{equation}
{}From this expression  the FSS relations can be derived.

{}From (\ref{thesis5.101}) and (\ref{thesis5.103}), if $t=0$
\begin{equation}
 W_l(0,\vec{H})
 =
 c_1^\prime
 (\ln{l})^{\frac{1}{3}}
 ({\vec{H}}^2)^{\frac{2}{3}}
\quad ,
\label{thesis7.21}
\end{equation}
and if $H = 0$
\begin{equation}
  W_l(t,\vec{0})
  \propto
  \left\{
         \begin{array}{ll}
          t^2
          \left[
                c_2^\prime (\ln{l})^{\frac{4-n}{n+8}} + {\rm{const.}}
          \right]
                                         &  {\rm if\ }  n \neq  4
                                                               \\
          t^2
          \left[
                c_3^\prime \ln{(\ln{l})} + {\rm{const.}}
          \right]
                                         & {\rm if \ }  n = 4
\quad .
               \end{array}
          \right.
\label{thesis7.16}
\end{equation}
Here $c_1^\prime$ and $c_2^\prime$  have the form
\begin{equation}
  {\rm{const.}}
  \left\{ 1 + O\left( \frac{\ln{(\ln{l})}}{\ln{l}}
  \right)  \right\}
\quad ,
\label{3rdc}
\end{equation}
while $c_3^\prime$ is
\begin{equation}
  {\rm{const.}}
  \left\{ 1 + O\left( \frac{1}{\ln{l}}
  \right)  \right\}
\quad .
\label{4thc}
\end{equation}

\section{FSS for $O(n)$ $\phi^4_4$ Theory}
\setcounter{equation}{0}

Let ${\vec{H}}$ define the first direction of internal space such that
$H_\alpha  =  H \delta_{1,\alpha}$.  From (\ref{thesis7.21}) the total
free energy at $t=0$ is   $c_1^\prime l^4 (\ln{l})^{1/3} H^{4/3}$. The
partition function is therefore
\begin{equation}
 Z_l(0,\vec{H})
 =
 Q\left(
 c_1^\prime l^4 (\ln{l})^{\frac{1}{3}} H^{\frac{4}{3}}
 \right)
\quad ,
\end{equation}
for  some  unknown  function $Q$.   At  a  Lee--Yang zero, $H_j$, this
vanishes.  Following \cite{IPZ} the FSS formula for this zero is found
to be
\begin{equation}
 H_j
 \sim
 l^{-3}
 (\ln{l})^{-\frac{1}{4}}
 \left\{
      1 + O\left(
                  \frac{\ln{(\ln{l})}}{\ln{l}}
           \right)
 \right\}
\quad .
\label{FSSLY}
\end{equation}
Similar considerations can be applied to the Fisher zeroes.  The  free
energy for vanishing $H$ is given by (\ref{thesis7.16}).  Setting  the
corresponding partition function to zero and solving  for   $t$  gives
the following FSS formula for the Fisher zeroes  in  four  dimensions;
\begin{equation}
 {t_j}^{-2}
 \sim
 \left\{
  \begin{array}{ll}
                     l^{4}
        [ c_2^\prime (\ln{l})^{\frac{4-n}{n+8}} +  {\rm{const.}} ]
                                         &  {\rm if\ }  n \neq  4
                                                              \\
                     l^{4}
                     [ c_3^\prime \ln{(\ln{l})} + {\rm{const.}}]
                                         & {\rm if \ }  n = 4
  \quad .
  \end{array}
 \right.
\label{FSSF}
\end{equation}
This  completes  the  list of finite--size formulae  for the partition
function   zeroes in four dimensions. The $n$--dependence   of the FSS
behaviour of the  Lee--Yang   zeroes   is found   in     the  additive
corrections to  (\ref{FSSLY}).

Knowledge of the scaling behaviour of the partition function zeroes is
equivalent  to knowing the scaling behaviour of the partition function
itself and of all derivable thermodynamic  quantities.   Writing   the
partition function as a product over its Lee--Yang zeroes,
\begin{equation}
 Z_l(t,\vec{H}) \propto \prod_j{\left( H-H_j \right)}
\label{old4.9}
\quad ,
\end{equation}
the magnetic susceptibility, given by the second derivative of the free
energy with respect to $H$, is
\begin{equation}
 \chi_l(t,\vec{H}) \propto \frac{1}{l^4}
 \sum_j{\frac{1}{\left(H-H_j\right)^2}}
\quad .
\end{equation}
The zero field susceptibility is then
\begin{equation}
 \chi_l \left( t,\vec{0} \right)
 \propto \frac{1}{l^4} \sum_j{ \frac{1}{H_j^2} }
\quad ,
\end{equation}
which from (\ref{FSSLY}) gives the FSS formula
\begin{equation}
 \chi_l(0,\vec{H}=\vec{0}) \propto l^2 (\ln{l})^{\frac{1}{2}}
 \left\{
        1 + O\left(
                   \frac{\ln{(\ln{l})}}{\ln{l}}
             \right)
 \right\}
\quad .
\label{fssforchiin4d}
\end{equation}
In terms of the Fisher zeroes (in the absence of an odd external
ordering field)
\begin{equation}
 Z_l(t,\vec{0}) \propto \prod_j{\left( t-t_j \right)}
\quad .
\end{equation}
The specific heat, given by the second derivative of the free energy
with respect to $t$, is
\begin{equation}
 C_l(t) = -\frac{1}{l^4}
 \sum_j{\frac{1}{\left(t-t_j\right)^2}}
\quad .
\end{equation}
At $t=0$, then
\begin{equation}
 C_l \left( t=0 \right)
 = -\frac{1}{l^4} \sum_j{ \frac{1}{t_j^2} }
\quad .
\end{equation}
In four dimensions (\ref{FSSF}) gives
\begin{equation}
 C_l(t=0)
 \sim
 \left\{
 \begin{array}{ll}
         c_2^\prime (\ln{l})^{\frac{4-n}{n+8}}
         +  {\rm{const.}}
                                         &  {\rm if\ } n \neq 4
                                                             \\
         c_3^\prime \ln{(\ln{l})} + {\rm{const.}}
                                         &  {\rm if\ } n = 4
 \quad .
 \end{array}
 \right.
\label{FSSCv}
\end{equation}

It has been known for a long time that FSS in the form (\ref{FSShyp})
breaks   down   in   four   dimensions. A modified version of the FSS
hypothesis, holding in and below four dimensions,     was proposed in
\cite{KeLa93}. This is
\begin{equation}
  \frac{P_l(0)}{P_\infty(t)}
  =
  f\left(
   \frac{\xi_l (0) }{\xi_\infty (t)}
  \right)
\quad .
\label{modFSShyp}
\end{equation}
Below   four   dimensions   where $\xi_l  \propto l$ this reduces  to
(\ref{FSShyp}).   In   four   dimensions   there exist multiplicative
logarithmic  corrections to the finite volume correlation length too.
To leading order this is \cite{Br82}
\begin{equation}
 \xi_\l (0) \sim  l(\ln{l})^{\frac{1}{4}}
\quad .
\end{equation}
This    modified   hypothesis   succeeds in recovering the above  FSS
formulae for the zero field susceptibility and the specific heat   in
four dimensions.

\section{Conclusions}
\setcounter{equation}{0}

Renormalization group techniques and perturbation theory have been  used
to  derive  the  finite--size  scaling  behaviour of the $O(n)$ $\phi^4$
theory in  four dimensions.

In  the four dimensional version  of  the  theory  there  appear certain
subtleties  not  present   below   four   dimensions.  The   IRFP of the
Callan--Symanzik function $B(g_R)$ moves to the origin as  the dimension
becomes  four  (the  perturbative  approach  predicts that the theory is
trivial). Secondly, in contrast to the $d<4$ dimensional case, the fixed
point is now  a double zero and this is responsible for  the  occurrence
of   logarithmic   corrections.   A  third  difference  comes  from  the
inhomogeneous term in the  RGE. The graph responsible for this   term is
not    divergent in less than   four dimensions where singular behaviour
comes from the homogeneous term. In four dimensions  the   inhomogeneous
term can also contribute to the leading scaling    behaviour.

The multiplicative logarithmic corrections to the leading mean field FSS
behaviour are, then, of primary interest. While the leading  logarithmic
corrections are dependent on the number of field components  $n$ for the
Fisher zeroes (and specific heat), they are $n$--independent in the case
of  the  Lee--Yang  zeroes (and the magnetic susceptibility). Any future
numerical  FSS  analysis of this $n$--dependence  should therefore be in
terms of Fisher zeroes or even thermodynamic functions.
\\

\noindent
{\bf Acknowledgement:}
The author would like to thank C.B. Lang for discussions and advice.
\newpage

\newpage


\end{document}